\documentclass[conference]{IEEEtran}
\IEEEoverridecommandlockouts
\usepackage{cite}
\usepackage{amsmath,amssymb,amsfonts}
\usepackage{algorithmic}
\usepackage{graphicx}
\usepackage{textcomp}
\usepackage{xcolor}
\usepackage{array}
\usepackage[hidelinks]{hyperref}
\usepackage[capitalize,noabbrev]{cleveref}
\usepackage{makecell}
\usepackage[T1]{fontenc}
\usepackage{booktabs}
\usepackage{multirow}
\usepackage[utf8]{inputenc}

\def\BibTeX{{\rm B\kern-.05em{\sc i\kern-.025em b}\kern-.08em
    T\kern-.1667em\lower.7ex\hbox{E}\kern-.125emX}}
\newcolumntype{M}[1]{>{\centering\arraybackslash}m{#1}}

\begin{document}

\title{Visual Set Program Synthesizer}

\author{Zehua Cheng$^{1,2}$, Wei Dai$^2$, Wenhu Zhang$^3$, Thomas Lukasiewicz$^{4,1}$, and Jiahao Sun$^2$\\
$^1$Department of Computer Science, University of Oxford\\
$^2$FLock.io\\
$^3$Department of Computer Science and Engineering, Hong Kong University of Science and Technology\\
$^4$Institute of Logic and Computation, TU Wien\\
\texttt{zehua.cheng@cs.ox.ac.uk}
}

\maketitle

\begin{abstract}

A user pointing their phone at a supermarket shelf and asking “Which soda has the least sugar?” poses a difficult challenge for current visual AI assistants. Such queries require not only object recognition, but explicit set-based reasoning such as filtering, comparison, and aggregation. Standard end-to-end MLLMs often fail at these tasks because they lack an explicit mechanism for compositional logic. We propose treating visual reasoning as Visual Program Synthesis, where the model first generates a symbolic program that is executed by a separate engine grounded in visual scenes. We also introduce Set-VQA, a new benchmark designed specifically for evaluating set-based visual reasoning. Experiments show that our approach significantly outperforms state-of-the-art baselines on complex reasoning tasks, producing more systematic and transparent behavior while substantially improving answer accuracy. These results demonstrate that program-driven reasoning provides a principled alternative to black-box visual–language inference.

\end{abstract}


\section{Introduction}

Modern Multimodal Large Language Models (MLLMs) have achieved strong progress in visual-language understanding, enabling systems to answer questions grounded in images~\cite{liu2023visual,achiam2023gpt}. However, they still struggle with queries that require explicit logical reasoning over sets of visual objects rather than simple recognition. For example, answering “Which soda on the shelf contains the least sugar?” requires identifying all sodas, retrieving the sugar content of each item, and selecting the minimum value. This type of set-based visual reasoning requires structured, step-wise operations over dynamically formed object collections, which goes beyond conventional Visual Question Answering.

Existing MLLMs typically address such problems through a single end-to-end inference process. The reasoning procedure remains implicit inside the model, which makes error diagnosis difficult and often leads to brittle behavior, such as incorrect filtering or comparison~\cite{biten2019scene}. Even tool-augmented systems still express reasoning in natural language form, making it challenging to verify whether intermediate steps are logically correct. As a result, current approaches lack both transparency and reliable control over the reasoning process.

\begin{figure}
    \centering
    \includegraphics[width=\linewidth]{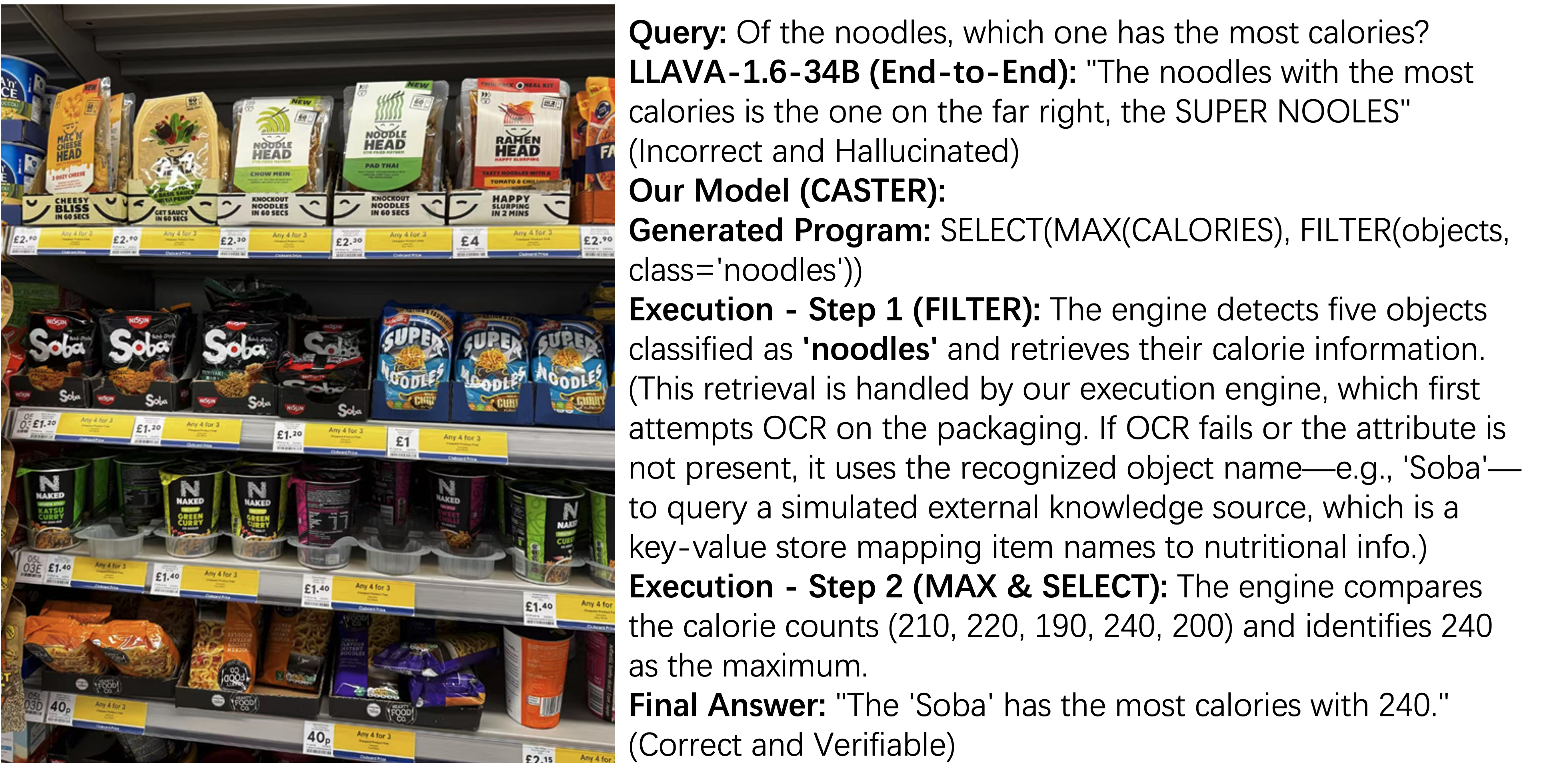}
    \caption{An Example of Visual Set Programmer with CASTER in Action Comparing with End-to-End Multi-modal LLM.}
    \label{fig:enterlabel}
\end{figure}

In this work, we formulate set-based visual reasoning as a Visual Program Synthesis problem. Rather than directly predicting an answer, the MLLM generates an executable symbolic program that represents the reasoning process. As is shown in Fig.~\ref{fig:enterlabel}, a separate execution engine then grounds the program in the visual scene and retrieves any necessary attributes, such as
\texttt{SELECT(MIN(sugar), FILTER(objects, class='soda'))}. This separation allows the model to reason through explicit set operations such as filtering, sorting, counting, and comparison. Compared with purely end-to-end modeling, this design makes the reasoning pipeline verifiable, since each intermediate operation produces a concrete output. It also enables more reliable error attribution, allowing us to distinguish perception failures from logical mistakes. It also provides interpretability, since each step can be inspected and evaluated individually.

Beyond improving task accuracy, this formulation also offers a unified perspective on multimodal reasoning, linking visual perception with symbolic computation. In practice, it enables models to handle increasingly complex queries without relying on brittle, implicit neural heuristics. We view this as a step toward building trustworthy and auditable visual AI assistants.
 
To further improve program generation quality, we introduce \textbf{Compositional Actor-Set Theoretic Reward (CASTER)}, a dense reward function that evaluates the correctness of intermediate program components rather than relying solely on final answer accuracy. This design provides richer learning signals and encourages the model to construct logically coherent programs, leading to more stable and interpretable reasoning behavior.

In summary, our contributions are as follows:
\begin{itemize}
    \item We formalize \textbf{set-based visual reasoning} as a critical but under-addressed challenge for MLLMs, requiring explicit compositional logic over visual object collections.
    \item We propose the \textbf{Visual Set Program Synthesizer (VSPS)}, a novel neuro-symbolic framework where an MLLM acts as a Visual Program Synthesizer, synthesizing and executing formal programs to answer visual queries reliably.
    \item We introduce the \textbf{Set-VQA} benchmark, a new dataset designed specifically to train and evaluate this capability, featuring egocentric images paired with set-based queries and their corresponding ground-truth programs that cover a wide range of logical compositions.
    \item Extensive experiments indicate our programmatic approach significantly outperforms state-of-the-art end-to-end MLLM baselines, dramatically reducing logical errors and improving final answer accuracy.
\end{itemize}

\section{Related Work}
\textbf{End-to-End Visual Reasoning with Implicit Structures}. Before the emergence of LLM-based models, compositional visual reasoning was typically addressed via monolithic deep nets that modeled relationships in an implicit manner. These models comprise Graph Convolutional Networks (GCNs) for constructing graphical relationships between visual entities and scene texts~\cite{biten2019scene,cheng2024affinity}, and other composite networks that combined individual deep nets, called "neural modules," to answer queries about images~\cite{gupta2023visual,yi2018neural}. Although these models can effectively represent correlations between entities, the compositional process is opaque and combined via a ``holistic" model. This compositional process is hardly precise enough for symbolic operations such as counting or filtering based on a condition, and is thus vulnerable to illogical errors. Our proposed Visual Set Program model ensures the compositional process is instead transparent and verifiable via the formal programs it generates.

\textbf{Visual Reasoning via Program Synthesis} We are related to the nascent line of work called neuro-symbolic visual reasoning, which conceives visual reasoning as a code synthesis problem. Here, a pre-trained LLM plays the role of a programmer who writes executable code to utilize specific vision models to respond to a question. Recent approaches such as VisProg\cite{gupta2023visual} and ViperGPT\cite{suris2023vipergpt} have showed that LLM can program code using pre-trained LLMs for complex and compositional visual problems in zero-shot or a few-shot setting. Recent work emphasized boosting the code generation process with robust hierarchical code synthesis models\cite{puadurean2023neural}.
Although we generalize on the same insight to break down inference with an executable program, there are three conceptual novelties in this work. 

\section{Methodologies}

\begin{figure*}[t]
    \centering
    \includegraphics[width=0.7\linewidth]{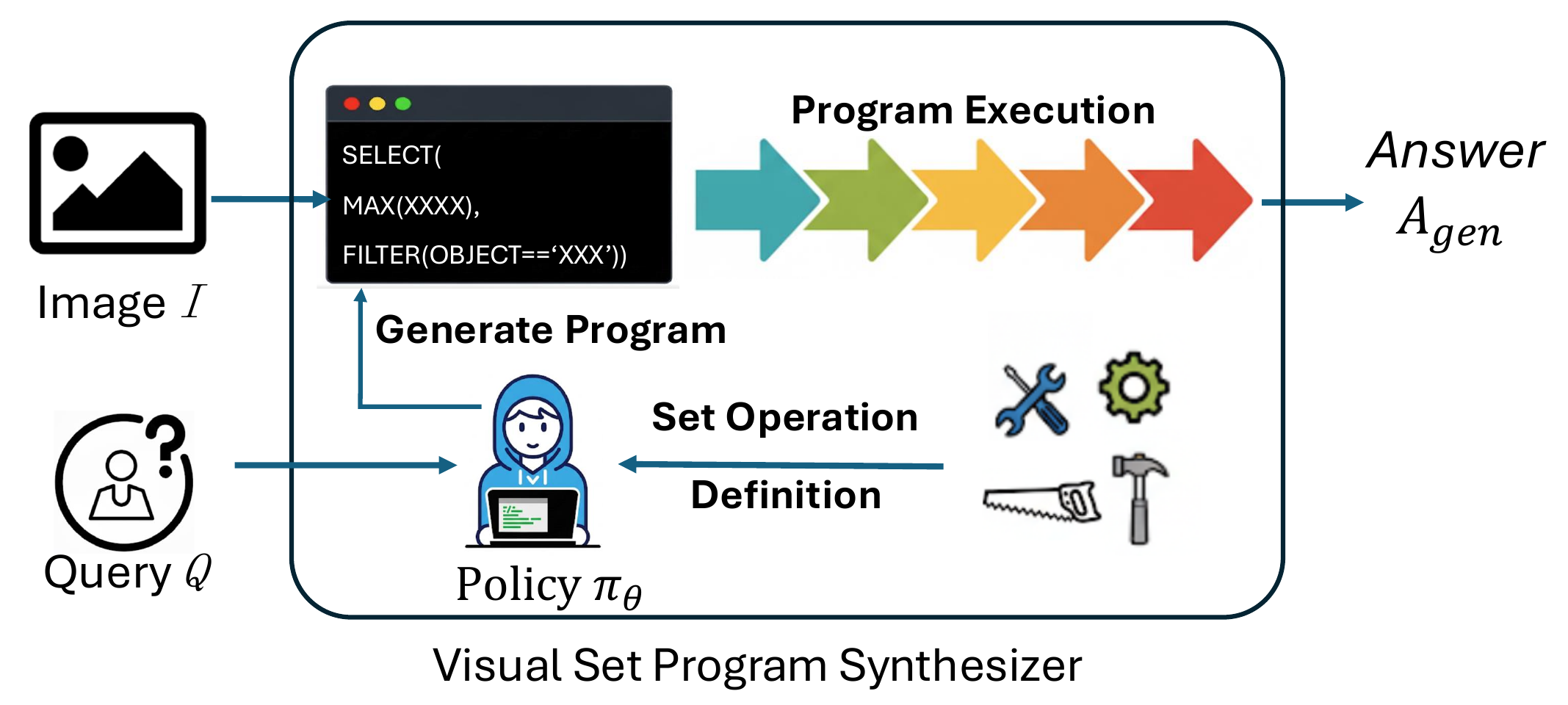}
    \vspace{-1em}
    \caption{Visual Set Program Synthesizer is a framework for solving complex visual queries by explicitly introducing a structured machine-readable program language. The MLLM generates a program, which is then run by the Program Execution Engine.
    This engine relies on a perception stack (e.g., object detection, OCR) and a knowledge base to ground the program's set operations (e.g., FILTER, SELECT) in the visual scene and retrieve necessary attributes.}
    \label{fig:method}
\end{figure*}

\subsection{Problem Formulation}
The central challenge is to train an MLLM to translate a multi-modal source—a visual cue $I$ and a natural language question $Q$—into a formal, executable code $P$. This executable code needs to express the compositional logic for querying. The goal is to develop a policy $\pi_\theta$ that generates a code $P$ for a given $\left(I, Q\right)$ such that executing it gives the correct answer $A_{gt}$. This is not mere child's play but demands immense expertise in programming language syntax and understanding natural language queries in terms of logical operations. To make this concrete, we define the program language and execution engine in Appendix~\ref{appx:def-plee}.

\subsection{Compositional Actor-Set Theoretic Reward (CASTER)}
The overall structure of proposed CASTER is presented in Fig.~\ref{fig:method}.
One such recent and immediately relevant baseline for programs with verifiable outcomes such as ML programs can be Reinforcement Learning with Verifiable Rewards (RLVR) ~\cite{mroueh2025reinforcement,wen2025reinforcement}. This involves optimizing MLLMs for successful completion of a task instead of relying on mere Supervised Fine-Tuning (SFT) to mimic correct programs in terms of syntax. RLVR is formulated with MLLMs modeled as policy networks that aim to produce programs by maximizing rewards based on correct answer evaluation.

The core of RLVR is a binary, verifiable reward function $\mathcal{R}$ that compares a generated answer $A_{gen}$ to a ground-truth answer $A_{gt}$:

\begin{equation}
    \mathcal{R}(A_{\text{gen}}, A_{\text{gt}}) = 
\begin{cases} 
1 & \text{if } A_{\text{gen}} = A_{\text{gt}} \\
0 & \text{otherwise}
\end{cases}
\end{equation}

The training objective is to maximize the expected reward $J(\theta)$ over the distribution of training data:
\begin{equation}
    J(\theta) = \mathbb{E}_{(I, Q, A_{\text{gt}}) \sim \mathcal{D}} \left[ \mathbb{E}_{P \sim \pi_{\theta}(\cdot|I, Q)} [R(\text{Exec}(P, I), A_{\text{gt}})] \right]
\end{equation}

This objective can be optimized using policy gradient methods like REINFORCE. The gradient of the objective function is given by:

\begin{equation}
    \nabla_{\theta} J(\theta) = \mathbb{E}_{P \sim \pi_{\theta}} [R(\text{Exec}(P, I), A_{\text{gt}}) \nabla_{\theta} \log \pi_{\theta}(P | I, Q)]
\end{equation}
To prevent such imbalances during training and mitigate the issue of forgetting what had been learned previously (catastrophic forgetting), trust region methods such as Proximal Policy Optimization (PPO)~\cite{schulman2017proximal}, Group Relative Policy Optimization (GRPO)~\cite{shao2024deepseekmath0}, and others have been explored. These methods add a penalty term involving the KL divergence to keep the new policy \(\pi_{\theta}\) close to a precomputed baseline policy \(\pi_{ref}\), such as from SFT optimization.

The RLVR baseline suffers from two major drawbacks that make it inefficient to learn complex compositional programs effectively:
(1) \textbf{Sparse Rewards}: The binary rewarding scheme does not provide any information about partially correct programs. Logically incorrect programs and logically correct but slightly flawed programs both receive the same reward of zero.
(2) \textbf{Credit Assignment Problem}: The global reward function doesn't provide any information about what part of the complex computation is wrong. The algorithm wouldn't be able to distinguish between errors in either the FILTER or MIN steps. 

To combat these challenges, we propose \textbf{Compositional Actor-Set Theoretic Reward (CASTER)}, a new neuro-symbolic learning technique wherein the training provides a dense and semantically rich reward signal. Crucially, CASTER does not learn from scratch; it explicitly optimizes an SFT-initialized policy, which provides the necessary syntactic priors for program generation. CASTER's fundamental novelty lies in framing the reward as a \emph{maximum weight bipartite matching} problem on Abstract Syntax Trees (ASTs), providing dense, semantic feedback via the Jaccard Index. This directly solves the sparse reward and credit assignment problems outlined above. Rather than focusing solely on answering the question correctly, CASTER rewards models for their compositional correctness in deriving the logical structure of their generated program. It accomplishes this by running all sub-programs and checking their generated sets of objects across all execution steps of these sub-programs against what would have been generated in the ground-truth program. In so doing, it provides learning feedback from sparse to dense rewards.

The CASTER framework extends the RLVR setup by parsing programs into Abstract Syntax Trees (ASTs) and computing a dense reward by comparing the semantic outputs of corresponding sub-programs. The algorithm proceeds by first generating a candidate program $P$ from the actor $\pi_\theta$ and parsing both $P$ and the ground-truth program $P_{gt}$ into their respective ASTs, $T$ and $T_{gt}$. Each node $n$ in an AST represents a sub-program $P_n$.

Next, we perform \textbf{semantic grounding of sub-programs}. A sub-program executor, \texttt{Exec\_sub}($P_n$, $I$), is used to evaluate each node $n \in N(T)$ and $n_{gt} \in N(T_{gt})$, where $N(\cdot)$ is the set of nodes in an AST. This execution yields intermediate sets of visually-grounded object identifiers, $O_n$ and $O_{n_gt}$, mapping symbolic program components to concrete visual evidence.

With these sets, we can compute a \textbf{structural reward via optimal alignment}. The semantic similarity between a generated sub-program node $n_i \in N(T)$ and a ground-truth node $n_j^{gt} \in N(T_{gt})$ is defined using the Jaccard Index $J$ of their resulting object sets:

\begin{equation}
    S_{ij} = J(O_{n_i}, O_{n_j^{gt}}) = \frac{|O_{n_i} \cap O_{n_j^{gt}}|}{|O_{n_i} \cup O_{n_j^{gt}}|}
    \label{eq:jaccard_idx}
\end{equation}

This calculation produces a similarity matrix $\mathbf{S}$ of size $|N(T)| \times |N(T_{gt})|$. The total reward for program $P$ is then elegantly framed as a \textbf{maximum weight bipartite matching} problem on the graph defined by the two node sets and the similarity matrix $\mathbf{S}$. Let $M$ be a matching between the nodes of the two ASTs. The optimal matching $M^*$ is the one that maximizes the total similarity:

\begin{equation}
    M^* = \underset{M}{\text{argmax}} \sum_{(n_i, n_j^{gt}) \in M} \mathbf{S}_{ij} \label{eq:bipartite}
\end{equation}

This problem can be solved efficiently using methods like the Hungarian algorithm. The dense reward for the program $P$ is the total weight of this optimal matching, which we define as a potential function $\Phi(P, P_{gt})$ that measures its maximal structural and semantic alignment with $P_{gt}$:

\begin{equation}
    R_{\text{CASTER}}(P) = \Phi(P, P_{gt}) = \sum_{(n_i, n_j^{gt}) \in M^*} S_{ij} \label{eq:norm}
\end{equation}

The objective of this matching-type reward is to encourage semantic robustness by granting rewards to a program if it matches syntactically but shares the same semantics as its sub-steps. Nevertheless, we observe that there are some issues associated with this form of reward. It fails to enforce type matching between matching nodes (for example, it should be permissible to match SORT nodes if their outputs match), and it fails to punish spurious nodes present in tree $T$.

This reward is no longer sparse; it grants partial credit for every correctly formulated and grounded sub-part of the program, directly addressing the credit assignment problem. This formulation is inspired by potential-based reward shaping~\cite{badnava2019new,forbes2024potential0based}, which effectively guides the agent through a complex state space.

Finally, we apply this dense reward to \textbf{policy optimization}. The CASTER algorithm is trained using GRPO, which works very effectively in this case because it doesn't require learning a separate critique neural network. The rewards to the batch of K programs $P_k$ sampled from $\pi_\theta$ are computed as their CASTER rewards $R_{\text{CASTER}}(P_k)$. These rewards can be normalized to get advantage estimates $A_k$, such as via z-scoring. The objective of GRPO optimizes the probability of programs whose relative rewards are greater in the batch:

\begin{equation}\small
    \mathcal{L}_{\text{GRPO}}(\theta) = - \mathbb{E} \left[ \sum_{k=1}^{K} A_k \cdot \log \pi_{\theta}(P_k | I, Q) + \beta \cdot D_{\text{KL}}(\pi_{\theta} || \pi_{\text{ref}}) \right]
\end{equation}
Note here that $\beta$ represents a regularization coefficient and $\pi_{ref}$ represents the SFT-initialized policy, helping to stabilize training. The neuro-symbolic feedback cycle developed in ~\cite{mileo2025towards} allows MLLMs to learn rules for logical decomposition and visual grounding in significantly more sample-efficient and robust ways than performed by the sparse-reward RLVR baseline.

\section{Experiments}
\subsection{Experimental Setup\label{sec:exp_setup}}
To evaluate how well our framework performs and generalizes to different scenarios of set reasoning, we use our newly developed models as benchmarks as well as two other sets of benchmarks from adapted VQA datasets. In Appendix~\ref{appx:hyperparameters}, we present hyperparameter values and additional information about baseline methods.

\begin{enumerate}
    \item \textbf{Set-VQA (Ours)}: Our primary benchmark for training and evaluation. It contains 110k (60k synthetic, 50k real-world) images (e.g., store shelves, desktops, street scenes) paired with complex natural language queries, ground-truth executable programs, and final answers. The queries are designed to cover a balanced distribution of set operations, including \texttt{FILTER}, \texttt{COUNT}, \texttt{SORT} (with \texttt{MIN}/\texttt{MAX}), \texttt{EXISTS}, and their compositions. The dataset is split into 80k training, 10k validation, and 20k test samples. The details of Set-VQA are presented in Appendix~\ref{appx:setvqa}.
    \item \textbf{GQA-Set}: To assess generalization to a new visual domain and question distribution, we created a challenging test split based on the GQA dataset~\cite{hudson2019gqa}, which is famous for its compositional questions and scene graphs. Five thousand questions were chosen such that they pertain to reasoning about sets of objects (for example, ``How many red things are to the left of the small sphere?''), and their ground truth answer programs were manually annotated.
    \item \textbf{PTR-Count}: To build the evaluation test split, we collected 3,000 samples from the Part-based visual reasoning (PTR)~\cite{hong2021ptr0}. The questions focus on counting object parts according to their attribute (for example, ``How many chairs have four legs?''). The test will determine the ability to reason about part-level aggregates of objects in three-dimensional space. The related programs were marked and validated simultaneously.
\end{enumerate}

To specifically evaluate compositional generalization (Table~\ref{tab:exp_generalization}), we ensured that the 80k training set explicitly excludes the following five complex, nested program structures, which appear only in the test set:
{\raggedright
\begin{itemize}
    \item Task 1: \texttt{COUNT(SORT(...))} (e.g., "Count the items after sorting by price").
    \item Task 2: \texttt{SORT(FILTER(...))} (e.g., "Sort the red items")
    \item Task 3: \texttt{COUNT(FILTER(objects, attribute != value))}(e.g., "How many items are not red?")
    \item Task 4: \texttt{SELECT(MAX(attr1),  SORT(attr2, ...))} (e.g., "Of the cheapest items, which has the most calories?")
    \item Task 5: {\small\texttt{COUNT(FILTER(objects, relation='left', ...))}} (e.g., "Count the cans to the left of the bottle")

\end{itemize}
}

\begin{table*}[t]

\caption{Main performance comparison on Set-VQA, GQA-Set, and PTR-Count benchmarks. PA and AA denote Program Accuracy and Final Answer Accuracy, respectively. Best results are in \textbf{bold}, second best in underline\label{tab:main_exp}}
    \centering
    \small
    \begin{tabular}{c|c|c|c|c}\toprule
    Model            & Type            & Set-VQA (PA/AA) & GQA-Set (PA/AA) & PTR-Count (PA/AA) \\\midrule
    \multicolumn{5}{c}{End-to-End Baselines}\\\hline
    Qwen2.5-VL-72B   & End-to-End      & N/A / 38.6\%      & N/A / 35.1\%      & N/A / 25.4\%        \\
    CogVLM-17B       & End-to-End      & N/A / 37.5\%      & N/A / 34.2\%      & N/A / 24.0\%        \\
    InternVL-2.5-26B & End-to-End      & N/A / 36.4\%      & N/A / 33.7\%      & N/A / 23.1\%        \\
    Qwen2.5-VL-32B   & End-to-End      & N/A / 36.1\%      & N/A / 32.8\%      & N/A / 23.9\%        \\
    LLaVA-1.6-72B    & End-to-End      & N/A / 34.8\%      & N/A / 31.5\%      & N/A / 22.4\%        \\
    LLaVA-1.6-34B    & End-to-End      & N/A / 31.2\%      & N/A / 28.5\%      & N/A / 19.8\%        \\
    Qwen2.5-VL-7B    & End-to-End      & N/A / 28.9\%      & N/A / 26.5\%      & N/A / 18.1\%        \\
    BLIP-2           & End-to-End      & N/A / 26.5\%      & N/A / 24.1\%      & N/A / 16.2\%        \\
    LLaVA-1.6-7B     & End-to-End      & N/A / 24.5\%      & N/A / 22.1\%      & N/A / 15.3\%        \\\midrule
    \multicolumn{5}{c}{Few-Shot Synthesizer Baselines}                                               \\\midrule
    ViperGPT         & Few-Shot Synth. & 22.1\% / 34.5\%   & 18.9\% / 29.8\%   & 14.3\% / 22.0\%     \\
    VisProg          & Few-Shot Synth. & 15.3\% / 25.8\%   & 11.2\% / 19.4\%   & 8.5\% / 14.2\%      \\\midrule
    \multicolumn{5}{c}{Visual Set Program Synthesizer (Qwen2.5-VL-7B Base)}                                 \\\midrule
    SFT (Ours)       & Ours            & 77.2\% / 73.5\%   & 70.1\% / 67.2\%   & 65.4\% / 61.3\%     \\
    RLVR (Ours)      & Ours            & 79.5\% / 86.1\%   & 73.1\% / 80.2\%   & 67.6\% / 74.5\%     \\
    CASTER (Ours)    & Ours            & \textbf{95.1\% / 93.8\%}   & \textbf{91.2\% / 89.9\%}   & \textbf{88.3\% / 87.1\%}     \\\midrule
    \multicolumn{5}{c}{Visual Set Program Synthesizer (LLaVA-1.6-7B Base)}                                  \\\midrule
    SFT (Ours)       & Ours            & 75.8\% / 71.3\%   & 68.2\% / 64.9\%   & 63.5\% / 59.1\%     \\
    RLVR (Ours)      & Ours            & 78.2\% / 84.6\%   & 71.5\% / 78.3\%   & 65.9\% / 72.4\%     \\
    CASTER (Ours)    & Ours            & \underline{94.3\% / 92.1\%}  & \underline{89.6\% / 87.5\%}   & \underline{86.2\% / 84.9\%}    \\\bottomrule
    \end{tabular}
\end{table*}

\subsection{Evaluation}
We use metrics that assess both the intermediate program and the final answer to provide a holistic evaluation.

\begin{enumerate}


    \item \textbf{Program Accuracy (PA)}: The percentage of generated programs that are an exact match to the ground-truth program string. This metric evaluates the synthesizer's ability to produce logically correct plans. A score of N/A is assigned to models that do not generate programs.

    \item \textbf{Final Answer Accuracy (AA)}: The percentage of generated answers that exactly match the ground-truth answer after execution. This is the primary measure of overall task success.

\end{enumerate}

\subsection{Main Results}



The Visual Set Program Synthesizer (VSPS) framework significantly outperforms all baseline methods, including both end-to-end models and few-shot program synthesizers. The best configuration (CASTER-Qwen) reaches 93.8\% Average Accuracy (AA) on Set-VQA, representing an absolute improvement of more than 55 points over the strongest monolithic baseline (Qwen2.5-VL-72B, 38.6\% AA). It also surpasses the most competitive tool-augmented baseline (51.2\% AA), demonstrating that simply granting models access to external perception or knowledge sources is insufficient. What truly matters is the ability to express reasoning explicitly in the form of an executable program, rather than relying on implicit neural inference.

Across both backbone architectures, the CASTER learning algorithm consistently yields superior performance. With the Qwen backbone, it improves Program Accuracy (PA) by more than 15 points (95.1\% vs. 79.5\%) and Final Answer Accuracy (FAA) by over 7 points (93.8\% vs. 86.1\%) relative to RLVR. Although RLVR achieves relatively high AA (86.1\%), this occurs despite its low PA (79.5

The framework demonstrates strong generality, achieving 93.8\% and 92.1\% AA with Qwen and LLaVA backbones, respectively. This suggests that the paradigm itself, rather than the choice of underlying language model, is the key driver of performance.

\textbf{Fairness of Baselines: Knowledge-Augmented Evaluation.}
A natural concern is whether the performance gap between CASTER and end-to-end baselines arises simply because our execution engine provides access to external knowledge (e.g., nutritional information from a knowledge base) that is unavailable to standard models. To ensure a strictly fair comparison, we conduct a knowledge-augmented evaluation where the exact same key-value external knowledge is injected directly into the text prompts of all end-to-end and few-shot baselines. Table~\ref{tab:augmented} reports results under both standard (St.) and augmented (Aug.) inference settings on a Set-VQA test subset.

Even with access to the same external knowledge, the strongest end-to-end model (Qwen2.5-VL-72B) improves only modestly, from 38.6\% to 42.1\% Overall AA. Critically, its performance on complex reasoning categories such as Extrema (31.2\%) and Exclusion (28.6\%) remains drastically below CASTER's 91.2\% and 90.1\%, respectively. This confirms that end-to-end models fail not because they lack information, but because they lack an explicit mechanism for compositional logic. The few-shot synthesizers (VisProg, ViperGPT) also improve under augmentation but still fall far short of CASTER, further demonstrating that reliable set-based reasoning requires structured program synthesis rather than unconstrained natural-language inference.

\begin{table*}[t]
\caption{Comparison of end-to-end models versus CASTER under standard (St.) and knowledge-augmented (Aug.) inference settings on the Set-VQA test subset. Augmented baselines receive the same external knowledge as CASTER's execution engine, injected directly into text prompts. VP = VisProg, VG = ViperGPT, CR = CASTER.\label{tab:augmented}}
\centering\small
\resizebox{\linewidth}{!}{
\begin{tabular}{c|c|c|ccccc}\toprule
Model & Params & Inf. Setting & Single-Step AA & Multi-Hop AA & Extrema AA & Exclusion AA & Overall AA \\\midrule
\multirow{2}{*}{Qwen2.5}  & \multirow{2}{*}{72B} & St.  & 46.2 & 31.5 & 29.8 & 25.1 & 38.6 \\
                          &                      & Aug. & 51.4 & 36.1 & 31.2 & 28.6 & 42.1 \\\hline
\multirow{2}{*}{LLaVA1.6} & \multirow{2}{*}{72B} & St.  & 42.1 & 28.6 & 24.5 & 21.0 & 34.8 \\
                          &                      & Aug. & 46.8 & 32.5 & 26.8 & 23.4 & 37.9 \\\hline
\multirow{2}{*}{Qwen2.5}  & \multirow{2}{*}{32B} & St.  & 43.8 & 29.4 & 27.1 & 22.8 & 36.1 \\
                          &                      & Aug. & 47.5 & 33.8 & 28.5 & 24.9 & 39.4 \\\hline
\multirow{2}{*}{LLaVA1.6} & \multirow{2}{*}{34B} & St.  & 38.5 & 24.1 & 20.4 & 18.5 & 31.2 \\
                          &                      & Aug. & 43.2 & 28.6 & 22.1 & 20.2 & 33.7 \\\hline
\multirow{2}{*}{Qwen2.5}   & \multirow{2}{*}{7B}  & St.  & 36.4 & 21.8 & 18.2 & 16.1 & 28.9 \\
                          &                      & Aug. & 41.5 & 25.3 & 20.0 & 18.3 & 31.5 \\\hline
\multirow{2}{*}{LLaVA1.6}  & \multirow{2}{*}{7B}  & St.  & 31.2 & 17.9 & 15.1 & 13.8 & 24.5 \\
                          &                      & Aug. & 36.8 & 21.4 & 16.5 & 15.0 & 26.8 \\\midrule
\multirow{2}{*}{VP (GPT-4o)}  & \multirow{2}{*}{N/A} & St.  & 38.2 & 22.5 & 19.5 & 18.2 & 25.8 \\
                              &                      & Aug. & 44.5 & 26.8 & 22.1 & 20.5 & 31.4 \\\hline
\multirow{2}{*}{VG (GPT-4o)}  & \multirow{2}{*}{N/A} & St.  & 46.8 & 29.1 & 24.8 & 22.6 & 34.5 \\
                              &                      & Aug. & 53.4 & 34.2 & 28.5 & 26.1 & 40.8 \\\midrule
\textbf{\makecell{CASTER\\ (Qwen2.5)}} & 7B & \makecell{Exec.\\ Engine} & \textbf{97.8} & \textbf{94.2} & \textbf{91.2} & \textbf{90.1} & \textbf{93.8} \\\bottomrule
\end{tabular}
}
\end{table*}

\begin{table*}[t]
\caption{Detailed Compositional Generalization Performance. We evaluate on increasingly complex compositions, including logical exclusion (\texttt{FILTER\_NOT}), multi-level comparisons (\texttt{MAX(ATTR, SORT)}) (Task\#4), and combined spatial reasoning (\texttt{COUNT(FILTER\_SPATIAL)}) (The Task\#5). Task\#1 refers to Unseen COUNT(SORT); Task\#2 refers to Unseen SORT(FILTER); The Task\#3 refers to Unseen Exclusion FILTER\_NOT.
}\label{tab:exp_generalization}
\centering\small
\resizebox{\linewidth}{!}{
\begin{tabular}{c|c|ccccc}\toprule
Model &
  Overall (AA) &
  Task\#1 (PA/AA) &
  Task\#2 (PA/AA) &
  Task\#3 (PA/AA) &
  Task\#4 (PA/AA) &
  Task\#5 (PA/AA) \\\midrule
CASTER (Qwen)  & 88.90\% & 90.5\% / 89.5\% & 91.2\% / 90.1\% & 89.5\% / 88.1\% & 87.2\% / 86.5\% & 85.1\% / 83.4\% \\
CASTER (LLaVA) & 87.50\% & 89.1\% / 88.3\% & 90.3\% / 89.2\% & 88.0\% / 86.8\% & 85.9\% / 84.9\% & 83.5\% / 81.9\% \\
RLVR (Qwen)    & 74.50\% & 78.2\% / 75.1\% & 79.9\% / 76.3\% & 71.3\% / 68.2\% & 65.0\% / 61.7\% & 63.8\% / 59.5\% \\
SFT (Qwen)     & 58.20\% & 61.3\% / 55.4\% & 63.8\% / 60.1\% & 55.1\% / 51.5\% & 49.8\% / 44.3\% & 47.2\% / 41.1\% \\\midrule
Qwen2.5-VL-72B      & 21.30\% & N/A / 19.8\%    & N/A / 22.5\%    & N/A / 15.1\%    & N/A / 9.8\%     & N/A / 11.2\%    \\
CogVLM-17B          & 20.10\% & N/A / 18.5\%    & N/A / 21.1\%    & N/A / 13.5\%    & N/A / 8.5\%     & N/A / 10.5\%    \\
LLAVA-1.6-34B       & 15.80\% & N/A / 14.2\%    & N/A / 16.3\%    & N/A / 10.2\%    & N/A / 6.1\%     & N/A / 7.3\%     \\\midrule
ViperGPT            & 24.10\% & 25.1\% / 22.7\% & 28.0\% / 25.8\% & 20.1\% / 17.6\% & 15.7\% / 12.0\% & 18.9\% / 15.2\% \\
VisProg             & 16.50\% & 17.3\% / 15.1\% & 19.2\% / 17.2\% & 13.9\% / 11.4\% & 10.1\% / 7.8\%  & 12.5\% / 9.9\% \\\bottomrule
\end{tabular}
}

\end{table*}

\subsection{Compositional Generalization}

To push the boundaries of compositional reasoning, we evaluate a “Compositional-Split” setting (defined in Section~\ref{sec:exp_setup}), which challenges models with not only basic nesting but also set exclusion logic, multi-level comparisons, and spatial reasoning.
The results are reported in Table~\ref{tab:exp_generalization}, using Program Accuracy (PA) and Answer Accuracy (AA) as evaluation metrics. End-to-end models do not generate programs, so PA is marked as N/A for these methods.

The CASTER model maintains high accuracy even on the most difficult compositional tasks. When faced with queries requiring logical exclusion (\texttt{FILTER\_NOT}) or multi-step comparative reasoning (\texttt{MAX(ATTR, SORT)}), CASTER's Answer Accuracy remains impressively high at 88.1\% and 86.5\%. This demonstrates that its understanding of program semantics is deep enough to handle fundamentally new logical structures. Its performance on spatial queries (83.4\%) further shows it can successfully integrate different reasoning modalities.

End-to-end models have shown only limited ability to carry out traditional compositional questions and have shown absolute failure to answer such complex queries. For instance, Qwen2.5-VL-72B shows only 9.8\% accuracy for answering ``Unseen Comparative'' questions, which is ten times lower than what CASTER achieves. This highlights the difficulty end-to-end architectures face when explicit multi-step logical reasoning is required.


The baseline Qwen-7B-Tool performance, which is far better than end-to-end models but still somehow suboptimal (28.7\% overall AA), shows that while access to information through tools is very helpful, answering complex questions involving several steps of logically correct reasoning in natural language syntax is not robust.

Despite their performance advantage over end-to-end models, SFT and RLVR models appear to be quite brittle in this case. The performance of SFT drops to 44.3\% for the comparative task because it has not seen such a complex syntactic form of a program. Although comparatively better, RLVR drops significantly to 61.7\% because it is not able to learn from its sparse reward function to make such complex decisions and results in hacked programs that go wrong in such complex situations. 

\begin{table}[t]
\caption{Ablation of CASTER reward components on Set-VQA test set.\label{tab:ablation}}
\centering
\small
\begin{tabular}{c|c|c}
\toprule
Model & PA (mean±std) & AA (mean±std) \\\midrule
CASTER (Full) &  \textbf{95.1(±0.4)\%}&\textbf{93.8(±0.3)\%} \\ \hline
w/ Type-Only Match   & 91.2(±0.7)\% & 89.5(±0.6)\% \\ \hline
w/ Binary Reward     & 85.4(±1.0)\% & 84.0(±1.1)\% \\ \hline
w/ Normalized Reward & 94.8(±0.5)\% & 93.5(±0.4)\% \\ \hline
RLVR & 79.5(±1.2)\% & 86.1(±0.9)\% \\\bottomrule
\end{tabular}
\end{table}

\subsection{Ablation Studies}

We conduct ablation experiments on Set-VQA (Table~\ref{tab:ablation}) to isolate the contribution of each component of the CASTER reward. We study four factors.
(1) Reward Density: We test whether partial credit based on Eq.~\ref{eq:jaccard_idx} is more effective than binary feedback. In the CASTER (Binary Reward) variant, the Jaccard score $S_{ij}$ is replaced with a binary value that equals 1 only when the intermediate sets $O_n$ and $O_{n_{gt}}$ are identical.
(2) Matching Strategy: We examine whether semantic bipartite matching in Eq.~\ref{eq:bipartite} is necessary by comparing against a stricter variant, CASTER (Type-Only Match), in which nodes can only match others with the same operator type (e.g., FILTER to FILTER).
(3) Reward Normalization: To test whether longer programs inflate rewards, CASTER (Normalized Reward) divides the final $R_{CASTER}$ by the number of ground-truth AST nodes $|N(T_{gt})|$.
(4) Reward Sparsity: RLVR is included as a baseline that uses a purely sparse 0/1 answer reward.

We observe that CASTER (Binary Reward) reduces PA by about 10\%, confirming that partial semantic similarity provides more informative learning signals than strict correctness alone. The Jaccard index therefore plays an important role in rewarding partially correct sub-programs, allowing the model to gradually refine its reasoning structure rather than failing catastrophically.

The CASTER (Type-Only Match) variant also yields a clear drop in performance. This indicates that flexible semantic alignment is beneficial, since it gives credit to logically equivalent programs even when they differ syntactically, rather than penalizing harmless structural variation. This shows that compositional reasoning benefits from semantic rather than purely structural supervision.

Finally, reward normalization has only negligible impact. This suggests that reward inflation due to program length is not a major factor in practice, and that the overall performance gains of CASTER do not depend on program complexity.

\subsection{Error Analysis}
To understand the remaining failure modes of CASTER, we conduct a fine-grained analysis over the full Set-VQA test set (20,000 samples) by comparing standard pipeline execution against an \emph{Oracle Executor} that uses ground-truth object detections, thereby isolating perception errors from logical errors. Table~\ref{tab:error_analysis} reports the results across 19 individual logic operations and five levels of query complexity (Depth 1 through Depth 5+).

Overall, CASTER achieves 95.1\% PA and 93.8\% AA under standard execution. Comparing against the Oracle Executor (which achieves 95.1\% AA), we observe a \textbf{Logic Gap of only 4.9\%} and a \textbf{Perception Gap of just 1.3\%}. This indicates that the vast majority of CASTER's remaining errors stem from program generation rather than perceptual failures. At the operation level, basic operations (FILTER by Color/Shape, COUNT, EXISTS) show near-perfect accuracy ($\geq$96.5\% AA), while more complex operations involving OCR-dependent attributes (FILTER Brand/OCR: 92.4\% AA) or deep nesting (Depth 5+: 85.1\% AA) present greater challenges. Notably, even at Depth 5+, the perception gap remains small (1.5\%), confirming that the execution engine's detection stack is not the bottleneck; rather, generating deeply nested programs is inherently harder for the synthesizer.

This analysis reveals that \emph{the bottleneck has shifted from logic to perception}: CASTER has effectively solved the compositional reasoning challenge.

\begin{table}[t]
\caption{Fine-grained error analysis of CASTER (Qwen2.5-VL-7B) on Set-VQA. Standard vs.\ Oracle Executor results isolate the perception gap across logic operations and query depths.\label{tab:error_analysis}}
\centering
\resizebox{0.5\textwidth}{!}{
\begin{tabular}{c|c|cc|cc}\toprule
Logic Operation & Depth & Std. PA & Std. AA & \makecell{Logic\\ Gap} & \makecell{Perc.\\ Gap}\\\midrule
FILTER (Color) & \multirow{7}{*}{1} & 98.1 & 96.8 & 1.9 & 1.3 \\
FILTER (Shape) & & 97.9 & 96.5 & 2.1 & 1.4 \\
FILTER (Texture) & & 97.0 & 95.2 & 3.0 & 1.8 \\
FILTER (Brand/OCR) & & 95.5 & 92.4 & 4.5 & 3.1 \\
COUNT (Basic) & & 98.5 & 97.1 & 1.5 & 1.4 \\
EXISTS (Basic) & & 99.1 & 98.2 & 0.9 & 0.9 \\
SELECT (Basic) & & 97.5 & 96.0 & 2.5 & 1.5 \\\hline
SORT (Spatial) & \multirow{4}{*}{2} & 95.8 & 94.5 & 4.2 & 1.3 \\
SORT (Numeric) & & 95.1 & 93.9 & 4.9 & 1.2 \\
MIN (Attribute) & & 95.3 & 94.1 & 4.7 & 1.2 \\
MAX (Attribute) & & 95.2 & 93.8 & 4.8 & 1.4 \\\hline
FILTER+COUNT & \multirow{3}{*}{3} & 96.2 & 95.0 & 3.8 & 1.2 \\
FILTER+SELECT & & 95.6 & 94.2 & 4.4 & 1.4 \\
SORT+MIN/MAX & & 93.1 & 91.8 & 6.9 & 1.3 \\\hline
FILTER+SORT & \multirow{2}{*}{4} & 92.8 & 91.5 & 7.2 & 1.3 \\
EXCLUSION (NOT) & & 91.5 & 90.1 & 8.5 & 1.4 \\\hline
FLT+SORT+MIN & \multirow{3}{*}{5+} & 89.9 & 88.5 & 10.1 & 1.4 \\
MULTI-FLT+COUNT & & 90.5 & 89.2 & 9.5 & 1.3 \\
DENSE NESTED & & 86.6 & 85.1 & 13.4 & 1.5 \\\midrule
\textbf{Overall} & \textbf{All} & \textbf{95.1} & \textbf{93.8} & \textbf{4.9} & \textbf{1.3} \\\bottomrule
\end{tabular}}
\end{table}

\section{Conclusion}

We addressed the limitations of current MLLMs in compositional set-based visual reasoning by proposing the Visual Set Program Synthesizer (VSPS) framework. The core of VSPS is the Compositional Actor-Set Theoretic Reward (CASTER), a dense, semantics-aware reward function that encourages models to generate logically correct and interpretable programs. Our experiments demonstrate that CASTER not only surpasses all baseline methods but also mitigates the ``reward hacking'' behavior commonly observed in sparse-reward RL.
Moreover, VSPS outperforms a fair, tool-augmented baseline by 42 points, even when end-to-end models are provided with the same external knowledge, highlighting the advantage of explicit program synthesis over unconstrained tool use. Ablation studies confirm the necessity of each component of the proposed reward design, while our fine-grained error analysis shows that the primary failure modes now arise from perception rather than logic, with a perception gap of only 1.3\% versus a logic gap of 4.9\%. This points to a clear direction for future work: improving the underlying visual detectors and OCR modules. We believe our framework provides a principled foundation for building trustworthy, transparent multi-modal reasoning systems.

\section*{Acknowledgment}
This research was funded in part by the Austrian Science Fund (FWF) 10.55776/COE12 and the AXA Research Fund.
\clearpage
\bibliographystyle{IEEEbib}
\bibliography{icme2026references}

\vspace{12pt}

\clearpage
\appendices

\section{Program Language and Execution Engine Definition\label{appx:def-plee}}

Program Language: The MLLM generates programs in a domain-specific language (DSL) composed of nested set operations. The core operators include:
\begin{itemize}
    \item \texttt{FILTER(objects, <condition>)}: Returns a subset of objects based on an attribute (e.g., class='soda', calories > 100).
    \item \texttt{SELECT(attribute, objects)}: Retrieves a specific attribute from a set of objects.
    \item \texttt{COUNT(objects)}: Returns the cardinality of a set.
    \item \texttt{MIN(attribute, objects) / MAX(attribute, objects)}: Find the object with the min/max attribute value.
    \item \texttt{SORT(attribute, objects)}: Sorts the set.
    \item \texttt{EXISTS(objects)}: Returns true if the set is non-empty.
\end{itemize}

Execution Engine ($Exec$): This module is responsible for executing a given program $P$ against the image $I$. It is a deterministic, non-learnable component that bridges the symbolic program to the visual world.

\begin{itemize}
    \item \textbf{Perception Stack}: We use YOLO-World~\cite{Cheng2024YOLOWorld}, a pre-trained object detector to identify all potential objects in the scene and PaddleOCR with Model Context Protocol~\cite{hou2025model} to extract any visible text.
    \item \textbf{Attribute Grounding}: When a program requires an attribute (e.g., calories), the engine first attempts to find it via OCR.
    \item \textbf{Knowledge Retrieval}: If an attribute is not visually present (like 'calories' for a 'noodle' packet), the engine uses the object's class name to query a simulated, external knowledge base (a simple key-value dictionary).
    \item \textbf{Sub-program Executor} ($Exec\_sub$): This function, crucial for our reward, executes a sub-program (a node in the AST) and returns the set of concrete object identifiers it produces.
\end{itemize}

To note that the use of this execution engine, particularly the simulated knowledge source, provides our model with information not available to standard end-to-end MLLMs. To ensure a fair comparison, we will also evaluate a tool-augmented baseline MLLM in Section 4 that has access to the same perception and retrieval functions, but must learn to call them via natural language prompts rather than a structured program.

\section{Experimental Setup and Hyperparameters \label{appx:hyperparameters}}
To evaluate the robustness and flexibility of our framework, we implement the visual set programmer using two different open-source base MLLMs for the synthesizer: a Qwen2.5-VL-7B model and a LLaVA-1.6-7B model. This allows us to assess our method's performance across different underlying model architectures. For training, all our framework variants are first initialized via SFT for 3 epochs. The RLVR and CASTER models are then trained for an additional 5 epochs using GRPO with a KL-divergence coefficient $\beta$ of $0.05$ and a learning rate of $1e^{-6}$ with AdamW Optimizer. All experiments were conducted on $8$ NVIDIA H100 GPUs.

We compare visual set programmer against three classes of state-of-the-art models.

\textbf{End-to-End MLLMs}: These models are prompted to answer questions directly in a single step. We evaluate several leading models:
\begin{itemize}
    \item Qwen2.-VL~\cite{Qwen2.5-VL} (7B, 32B, 72B):  A family of state-of-the-art MLLMs known for their strong performance across a wide range of vision-language benchmarks. We include multiple sizes to test whether raw parameter scale can overcome the architectural deficit for set-based reasoning.
    \item CogVLM-17B~\cite{wang2023cogvlm}:  A powerful MLLM that features a visual expert module in its attention and FFN layers, enabling deep fusion of visual and textual information at multiple layers of the model, as opposed to only at the input layer.
    \item LLaVA-1.6 (7B, 34B, 72B)~\cite{liu2024llavanext}: A prominent and widely-used MLLM family that pioneered the use of GPT-generated data for visual instruction tuning. Its architecture connects a vision encoder to an LLM with a simple projection layer, making it a crucial baseline for instruction-following capabilities. We test multiple scales, including a hypothetical 72B variant, to assess its scaling properties on our task.
    \item BLIP-2~\cite{li2023blip}: A foundational MLLM that introduced the Q-Former, a lightweight bridge between a frozen image encoder and a frozen LLM. This parameter-efficient approach makes it a key architectural predecessor to many modern MLLMs.
\end{itemize}

\textbf{Few-Shot Program Synthesizers}: These models generate programs in a few-shot, prompting-based manner without task-specific training.
\begin{itemize}
    \item VisProg~\cite{gupta2023visual}: A foundational neuro-symbolic framework that generates Python code by composing visual modules based on in-context examples.
    \item ViperGPT~\cite{suris2023vipergpt}: An advanced framework that also generates Python programs to answer visual queries, notable for its ability to generate its own in-context examples to reduce manual prompt engineering.
\end{itemize}

\section{Set-VQA\label{appx:setvqa}}
The Set-VQA benchmark was designed to be a comprehensive test of set-based visual reasoning. To achieve this, it was built using a two-pronged approach to create its 110,000 examples:

The synthetic portion, consisting of 60,000 examples, was generated using a program-first methodology. This involved automatically generating complex, nested programs from a formal grammar. Then, a rendering engine, similar to CLEVR, created 3D scenes with objects that specifically matched the logic of each program. These formal programs were then translated into templated questions and finally paraphrased by human annotators to ensure natural language diversity.

The second portion consists of 50,000 real-world examples. This process began by sourcing egocentric images, such as photos of store shelves or desktops, which are rich with objects. Human annotators then wrote complex, set-based questions for these images. Following this, expert annotators, who were trained in the system's specific program language, wrote the corresponding ground-truth executable programs and verified the final answers. This combined approach ensures the benchmark is both logically complex (from the synthetic data) and visually realistic (from the real-world data).

Each data point in Set-VQA contains four key pieces of information. As shown in the example data point, this includes an image\_id like ``real\_shelf\_0451.jpg," a natural language query such as "Of the drinks on the top shelf, which one is the cheapest?", the corresponding ground-truth program which is ``SELECT(MIN(PRICE), FILTER(FILTER(objects, class='drink'), relation='on\_top\_shelf'))", and finally the correct final\_answer, which in this case is ``Spring Water". This structure is what allows the system to be trained and evaluated on its logical, programmatic reasoning.

\end{document}